\documentclass[12pt]{article}
\usepackage{a4wide}
\usepackage{latexsym}

\usepackage{pslatex}
\usepackage[latin1]{inputenc}
\usepackage[T1]{fontenc}
\usepackage{axodraw}
\usepackage{cite}

\def\bq{\begin{eqnarray}}
\def\eq{\end{eqnarray}}

\def\eps{\varepsilon}

\newtheorem{theorem}{Theorem}[section]

\begin{document}

\thispagestyle{empty}

\begin{flushright}
  MZ-TH/03-11 \\
  UPRF-2003-16
\\ 
\end{flushright}

\vspace{1.5cm}

\begin{center}
  {\Large\bf The massless two-loop two-point function\\}
  \vspace{1cm}
  {\large Isabella Bierenbaum$^a$ and Stefan Weinzierl$^b$}\\
  \vspace{1cm}
  $^a${\small {\em Institut f{\"u}r Physik,
  Universit{\"a}t Mainz}}\\
  {\small {\em 55099 Mainz, Germany}}\\[2mm]
  $^b${\small {\em Dipartimento di Fisica, Universit\`a di Parma,\\
       INFN Gruppo Collegato di Parma, 43100 Parma, Italy}
  } \\
\end{center}

\vspace{2cm}

\begin{abstract}\noindent
  {%
    We consider the massless two-loop two-point function 
    with arbitrary powers of the propagators
    and derive a representation, from which we can obtain the Laurent expansion
    to any desired order in the dimensional regularization parameter $\eps$.
    As a side product, we show that in the Laurent expansion of the two-loop integral
    only rational numbers and multiple zeta values occur.
    Our method of calculation obtains the two-loop integral as a convolution product
    of two primitive one-loop integrals.
    We comment on the generalization of this product structure to higher loop integrals.
   }
\end{abstract}

\vspace*{\fill}

\newpage

\section{Introduction}

Quantitative predictions from perturbation theory are crucially linked with
our ability to calculate loop integrals.
An object of extensive study has been the master two-loop two-point function,
with massless internal propagators but arbitrary powers of the propagators.
The name ``the master two-loop two-point function'' is justified, since all other
two-loop two-point topologies can be obtained from this one.
Allowing arbitrary powers of the propagators is important for three- or four-loop
calculations, where the integration over self-energy insertions on the propagators
shifts the power of the propagator from unity to $1+\eps$.
(As usual we work within dimensional regularization with $D=4-2\eps$.)

In this paper we consider the massless two-loop two-point function.
This integral is not only of practical importance from a phenomenological perspective,
but received also quite some interest from number theorists.
As far as the phenomenological side is concerned, this integral is implemented into
the {M{\sc incer}} package \cite{Gorishnii:1989gt,Larin:1991fz} and used for example
for the calculation of the total hadronic cross-section in electron-positron annihilation.
From the number theoretical perspective there has always been the open question which
types of (transcendental) numbers appear in the $\eps$-expansion of this integral.
From explicit calculations it is known, that in the lowest orders multiple zeta values
occur.

The history for the calculation of this integral dates back, to the best of our
knowledge, to 1980, when the $\eps^0$-coefficient for a specific (non-trivial)
combination of powers of propagators was calculated using the Gegenbauer polynomial
x-space technique \cite{Chetyrkin:1980pr}.
In 1984 Kazakov obtained for all powers of the propagators of the form $\nu_j=1+a_j \eps$
the result up to $\eps^3$, a year later the $\eps^4$-term followed
\cite{Kazakov:1983km,Kazakov:1984ns,Kazakov:1985pk}.
The momentum of pushing the $\eps$-expansion of this integral further was then
taken up by Broadhurst and collaborators: In 1986 the $\eps^5$-term was calculated,
in 1988 followed the $\eps^6$ term \cite{Broadhurst:1986bx,Barfoot:1988kg}.
In the mean time the original Gegenbauer technique had been refined \cite{Kotikov:1996cw}
and a representation of the integral for a specific combination of the powers of the propagators
in terms of hypergeometric ${}_3F_2$-functions with unit arguments was found
\cite{Kotikov:1996cw,Broadhurst:1997ur}.
The year 1996 brought an advance of two further terms in the expansion ($\eps^7$ and $\eps^8$)
\cite{Broadhurst:1997ur}.
Finally, last year the $\eps^9$-term was announced \cite{Broadhurst:2002gb}.
The calculation of most of these terms relied on symmetry properties of the two-loop integral.
It is known that this two-loop integral has the symmetry group $Z_2 \times S_6$
\cite{Gorishnii:1985te,Broadhurst:1986bx,Barfoot:1988kg}, 
which is of order $1440$, e.g. there are 1440 symmetry relations.
The calculations cited above exploited the fact that the symmetry properties allow to reconstruct
the result of the expansion up to order $\eps^9$ for powers of the propagators
of the form $\nu_j=1+a_j\eps$ from the result of the expansion, where two adjacent 
propagators occur with unit power.
However, it is known that beyond this order the symmetry relations are not sufficient to determine
the solution \cite{Broadhurst:1997ur}.
In view of the last point, 
the calculation of this two-loop integral cannot be considered
to be solved to a satisfactory level. We quote from the latest publication of 
Broadhurst \cite{Broadhurst:2002gb}:

{\it
``It is one of the many scandals of our limited understanding of the
analytical content of perturbative quantum field theory 
that, despite many years of intense effort, we still do not
know whether multiple zeta values suffice for even the Taylor expansion of the
two--loop integral.''
}

In this paper we proof that multiple zeta values are sufficient 
(theorem \ref{TheBierenbaumWeinzierlTheorem}).
This is one of the main results of this paper and solves a long standing open problem.
We also show how to obtain the $\eps$-expansion to arbitrary order for arbitrary powers of
the propagators by calculating this integral with the help of a new method.
For an introduction to calculational techniques for multi-loop
integrals we refer to \cite{Smirnov:2002kq,Weinzierl:2003jx,Grozin:2003ak}.

The second important result of this paper is the method we employed for the calculation
of the two-loop integral: We obtain the two-loop integral as a convolution product of two
primitive one-loop integrals. The convolution product can be evaluated in terms of nested sums
\cite{Moch:2001zr,Weinzierl:2002hv},
and the $\eps$-expansion of the original integral is obtained from the $\eps$-expansion of these
sums.
Generalizations of this product structure can be useful for the extension of exisiting
packages for two-point functions (like {M{\sc incer}} \cite{Gorishnii:1989gt,Larin:1991fz}) 
to four loops.
In this paper we only briefly comment on this possibility and focus on the two-loop two-point
function. 
In view of later applications for the calculations of higher orders we show that our method
for the calculation can be implemented efficiently on a computer.

This paper is organized as follows:
In sect. \ref{sect:known} we define the two-loop integral and
summarize known facts about this integral.
In sect. \ref{sect:prod} we show that the two-loop integral can be written as a 
convolution product of two one-loop integrals.
We also discuss the factorization for three-loop two-point functions.
In sect. \ref{sect:eval} we use this product structure to obtain the $\eps$-expansion
for the two-loop integral.
In sect. \ref{sect:checks} we report on the implementation 
of our formulae into a symbolic computer code.
Sect. \ref{sect:concl} contains a summary and our conclusions. 
An appendix collects some useful formulae for integral transformations.

\section{Review of known results for the two-loop integral}
\label{sect:known}

The object of investigation is the following five-propagator integral
\bq
\label{objofinvest}
\lefteqn{
\hat{I}^{(2,5)}(m-\eps,\nu_1,\nu_2,\nu_3,\nu_4,\nu_5)
 = } \nonumber \\
 & &
 c_\Gamma^{-2}
 \left( -p^2 \right)^{\nu_{12345}-2m+2\eps} 
 \int \frac{d^Dk_1}{i \pi^{D/2}}
 \int \frac{d^Dk_2}{i \pi^{D/2}}
  \frac{1}{ \left(-k_1^2\right)^{\nu_1}
            \left(-k_2^2\right)^{\nu_2}
            \left(-k_3^2\right)^{\nu_3}
            \left(-k_4^2\right)^{\nu_4}
            \left(-k_5^2\right)^{\nu_5}
          },
 \nonumber \\
\eq
where $k_3=k_2-p$, $k_4=k_1-p$, $k_5=k_2-k_1$, $D=2m-2\eps$ and
\bq
c_\Gamma & = & \frac{\Gamma(1+\eps)\Gamma(1-\eps)^2}{\Gamma(1-2\eps)}.
\eq
The superscripts for an integral $\hat{I}^{(l,n)}$ indicate the number of loops $l$ and
the number of propagators $n$.
The prefactor in front of the integral is inserted for later convenience.
It ensures that $\hat{I}^{(2,5)}$ is independent of $(-p^2)$ and avoids a proliferation
of Euler's constant $\gamma_E$.
The corresponding Feynman diagram for this two-loop integral is shown on the l.h.s of 
fig. \ref{fig1}.
It is for our purpose sufficient to assume that the exponents $\nu_j$ are of the form
\bq
\label{ourparametrization}
\nu_j & = & n_j + a_j \eps,
\eq
where the $n_j$ are positive integers and the $a_j$ are non-negative real numbers.
The $\eps$-dependence in the powers of the propagators arises from higher loop
integrals by integrating out simple one-loop two-point insertions into the propagators
of eq. (\ref{objofinvest}).

Integration-by-part identities \cite{'tHooft:1972fi,Tkachov:1981wb,Chetyrkin:1981qh}
relate integrals with different powers of the propagators.
For example, from the triangle rule \cite{vanRitbergen:1999fi}
we obtain the identities
\bq
\label{ibp}
\left[
\left( D - \nu_{235} - \nu_5 \right)
 + \nu_2 {\bf 2}^+ \left( {\bf 1}^- - {\bf 5}^- \right)
 + \nu_3 {\bf 3}^+ \left( {\bf 4}^- - {\bf 5}^- \right)
\right]
\hat{I}^{(2,5)} & = & 0,
\nonumber \\
\left[
\left( D - \nu_{145} - \nu_5 \right)
 + \nu_1 {\bf 1}^+ \left( {\bf 2}^- - {\bf 5}^- \right)
 + \nu_4 {\bf 4}^+ \left( {\bf 3}^- - {\bf 5}^- \right)
\right]
\hat{I}^{(2,5)} & = & 0.
\eq
Here, the operators ${\bf i}^+$ and ${\bf i}^-$ raise, respectively lower, 
the power of propagator $i$ by one.
The integration-by-part relations can be used to relate the integral 
where all propagators occur to a positive integer power to simpler
topologies, e.g. where one propagator is eliminated.

The integral in eq.(\ref{objofinvest}) has the obvious symmetries
\bq
\label{obvioussymmetry}
\left( \nu_1, \nu_2, \nu_3, \nu_4 \right)
 & \rightarrow &
\left( \nu_2, \nu_1, \nu_4, \nu_3 \right),
 \nonumber \\
\left( \nu_1, \nu_2, \nu_3, \nu_4 \right)
 & \rightarrow &
\left( \nu_4, \nu_3, \nu_2, \nu_1 \right).
\eq
However, there are more symmetries, which relate the integral to itself, up to prefactors
of products of Gamma functions.
To discuss the symmetry properties of eq. (\ref{objofinvest}) it is convenient
to introduce the function $F(\nu_0,\nu_1,\nu_2,\nu_3,\nu_4,\nu_5)$ related to
$\hat{I}^{(2,5)}$ by
\bq
\lefteqn{
\hat{I}^{(2,5)}(\nu_0,\nu_1,\nu_2,\nu_3,\nu_4,\nu_5)
 = }
\nonumber \\
& &
 \frac{\Gamma(2\nu_0-3)^2}{(2\nu_0-3) \Gamma(3-\nu_0)^2 \Gamma(\nu_0-1)^6}
 \left[
        \prod\limits_{j=1}^{10} \frac{\Gamma(\nu_0-\nu_j)}{\Gamma(\nu_j)}
 \right]^{1/2}
 F(\nu_0,\nu_1,\nu_2,\nu_3,\nu_4,\nu_5),
\eq
where
\bq
\nu_6 & = & 3 \nu_0 - \nu_{12345}
\eq
and 
\bq
\nu_7 = 2 \nu_0 - \nu_{235},
 & &
\nu_9 = \nu_{345} - \nu_0,
 \nonumber \\
\nu_8 = 2 \nu_0 - \nu_{145},
 & &
\nu_{(10)} = \nu_{125} - \nu_0.
\eq
Here and in the following we will use the short-hand notation like 
$\nu_{ijk} = \nu_i + \nu_j + \nu_k$ to denote sums of indices if $\{i,j,k\} \in \{1,2,3,4,5\}$.
The function $F(\nu_0,\nu_1,\nu_2,\nu_3,\nu_4,\nu_5)$ is invariant under the symmetry group $Z_2 \times S_6$ \cite{Barfoot:1988kg}.
The symmetric group $S_6$ is generated by the six-cycle
\bq
\label{sixcycle}
\left( \nu_0,\nu_1,\nu_2,\nu_3,\nu_4,\nu_5 \right)
 & \rightarrow &
\left( \nu_0,\nu_2,\nu_5,\nu_4,3\nu_0-\nu_{12345},\nu_3 \right)
\eq
and the transposition
\bq
\label{transposition}
\left( \nu_0,\nu_1,\nu_2,\nu_3,\nu_4,\nu_5 \right)
 & \rightarrow &
\left( \nu_0, -\nu_0+\nu_{145}, \nu_2, -\nu_0+\nu_{345}, \nu_0-\nu_5, \nu_0-\nu_4 \right).
\eq
The group $Z_2$ is generated by the reflection
\bq
\label{reflection}
\left( \nu_0,\nu_1,\nu_2,\nu_3,\nu_4,\nu_5 \right)
 & \rightarrow &
\left( \nu_0,\nu_0-\nu_1,\nu_0-\nu_2,\nu_0-\nu_3,\nu_0-\nu_4,\nu_0-\nu_5 \right).
\eq
Note that in general the generators of the symmetry group do not
conserve the positivity $a_j \ge 0$ in the parameterization 
of eq. (\ref{ourparametrization}).

\section{The product structure of the two-loop integral}
\label{sect:prod}

In this section we show that the two-loop integral in eq. (\ref{objofinvest})
can be written as a (convolution) product of two one-loop integrals.
To this aim we define 
the following two one-loop integrals:
\bq
\hat{I}^{(1,2)}(m-\eps,\nu_1,\nu_4)
 & = & 
 c_\Gamma^{-1}
 \left( -p^2 \right)^{\nu_{14}-m+\eps} 
 \int \frac{d^Dk_1}{i \pi^{D/2}}
  \frac{1}{ \left(-k_1^2\right)^{\nu_1}
            \left(-k_4^2\right)^{\nu_4}
          },
 \nonumber \\
I^{(1,3)}(m-\eps,\nu_2,\nu_3,\nu_5; x, y)
 & = & 
 c_\Gamma^{-1}
 \left( -p^2 \right)^{\nu_{235}-m+\eps} 
 \int \frac{d^Dk_2}{i \pi^{D/2}}
  \frac{1}{ 
            \left(-k_2^2\right)^{\nu_2}
            \left(-k_3^2\right)^{\nu_3}
            \left(-k_5^2\right)^{\nu_5}
          },
\eq
where $x=(-p^2)/(-k_1^2)$, $y=(-p^2)/(-k_4^2)$.
Note that the integral $I^{(1,3)}$ depends on the kinematic variables
$-p^2$, $-k_1^2$ and $-k_4^2$ only through the dimensionless ratios
$x$ and $y$.
The integral $\hat{I}^{(1,2)}$ is easily computed as
\bq
\label{oneloopbubble}
\hat{I}^{(1,2)}(m-\eps,\nu_1,\nu_4)
 & = & 
 c_\Gamma^{-1}
 \frac{\Gamma(-m+\eps+\nu_{14})}{\Gamma(\nu_1) \Gamma(\nu_4)}
 \frac{\Gamma(m-\eps-\nu_1) \Gamma(m-\eps-\nu_4)}{\Gamma(2m-2\eps-\nu_{14})}.
\eq
The one-loop triangle integral $I^{(1,3)}$ can be written as a double 
Mellin-Barnes representation:
\bq
\label{MellinBarnes}
I^{(1,3)}(m-\eps,\nu_2,\nu_3,\nu_5; x, y)
 & = &
 \frac{1}{(2 \pi i)^2}
 \int\limits_{\gamma_1-i \infty}^{\gamma_1+i \infty} d\sigma 
 \int\limits_{\gamma_2-i \infty}^{\gamma_2+i \infty} d\tau \;
 y^{-\sigma} x^{-\tau} \;
 \hat{I}^{(1,3)}(m-\eps,\nu_2,\nu_3,\nu_5; \tau, \sigma),
 \nonumber \\
\eq
where the function $\hat{I}^{(1,3)}(m-\eps,\nu_2,\nu_3,\nu_5; \tau, \sigma)$
is given by
\bq
\hat{I}^{(1,3)}(m-\eps,\nu_2,\nu_3,\nu_5; \tau, \sigma)
 & = &
 c_\Gamma^{-1}
 \frac{1}{\Gamma(\nu_2)\Gamma(\nu_3)\Gamma(\nu_5)\Gamma(2m-2\eps-\nu_{235})}
\nonumber \\
& &
 \times
 \Gamma(-\sigma) \Gamma(-\sigma+m-\eps-\nu_{35})
 \Gamma(-\tau) \Gamma(-\tau+m-\eps-\nu_{25})
 \nonumber \\
& &
 \times
 \Gamma(\sigma+\tau-m+\eps+\nu_{235}) \Gamma(\sigma+\tau+\nu_5).
\eq
The integration contours are parallel to the imaginary axis, with indentations,
if necessary, to separate the ``UV''-poles 
($\Gamma(-\sigma+...)$, $\Gamma(-\tau+...)$, $\Gamma(-\sigma-\tau+...)$) 
from the ``IR''-poles 
($\Gamma(\sigma+...)$, $\Gamma(\tau+...)$, $\Gamma(\sigma+\tau+...)$). 
It should be noted that the function 
$\hat{I}^{(1,3)}(m-\eps,\nu_2,\nu_3,\nu_5; \tau, \sigma)$
is the double Mellin transform in $x$ and $y$ of the original integral
$I^{(1,3)}(m-\eps,\nu_2,\nu_3,\nu_5; x, y)$:
\bq
\hat{I}^{(1,3)}(m-\eps,\nu_2,\nu_3,\nu_5; \tau, \sigma) & = & 
 \int\limits_0^\infty dx 
 \int\limits_0^\infty dy \;
 x^{\tau - 1} y^{\sigma -1} \;
 I^{(1,3)}(m-\eps,\nu_2,\nu_3,\nu_5; x, y).
\eq
From eq. (\ref{oneloopbubble}) and eq. (\ref{MellinBarnes}) 
one obtains the two-loop integral as
\bq
\label{convolutionprod}
\lefteqn{
\hat{I}^{(2,5)}(m-\eps,\nu_1,\nu_2,\nu_3,\nu_4,\nu_5)
 = } \nonumber \\
 & &
 \frac{1}{(2 \pi i)^2}
 \int\limits_{\gamma_1-i \infty}^{\gamma_1+i \infty} d\sigma 
 \int\limits_{\gamma_2-i \infty}^{\gamma_2+i \infty} d\tau \;
  \hat{I}^{(1,2)}(m-\eps,\nu_1-\tau,\nu_4-\sigma)
 \hat{I}^{(1,3)}(m-\eps,\nu_2,\nu_3,\nu_5; \tau, \sigma).
\eq
In eq. (\ref{convolutionprod})
the two-loop integral is obtained as a (double) convolution product
of two one-loop integrals.
\begin{figure}
\begin{center}
\begin{eqnarray*}
\begin{picture}(100,40)(20,45)
\Vertex(50,20){2}
\Vertex(50,80){2}
\Vertex(20,50){2}
\Vertex(80,50){2}
\Line(20,50)(50,80)
\Line(50,20)(20,50)
\Line(50,80)(80,50)
\Line(80,50)(50,20)
\Line(50,20)(50,80)
\Line(80,50)(100,50)
\Line(0,50)(20,50)
\Text(105,50)[l]{$p$}
\Text(32,65)[br]{$\nu_1$}
\Text(68,65)[bl]{$\nu_2$}
\Text(68,35)[tl]{$\nu_3$}
\Text(35,35)[tr]{$\nu_4$}
\Text(53,50)[l]{$\nu_5$}
\end{picture}
& = & 
\begin{picture}(120,40)(0,45)
\Vertex(25,50){2}
\CArc(50,50)(25,0,360)
\Line(75,50)(95,50)
\Line(5,50)(25,50)
\Text(100,50)[l]{$p$}
\Text(32,68)[br]{$\nu_1$}
\Text(32,32)[tr]{$\nu_4$}
\GCirc(75,50){5}{0.5}
\end{picture}
\star
\begin{picture}(100,40)(-10,45)
\Vertex(20,20){2}
\Vertex(20,80){2}
\Vertex(60,50){2}
\Line(20,20)(20,80)
\Line(20,80)(60,50)
\Line(60,50)(20,20)
\Line(60,50)(80,50)
\Line(0,20)(20,20)
\Line(0,80)(20,80)
\Text(85,50)[l]{$p$}
\Text(45,65)[bl]{$\nu_2$}
\Text(45,35)[tl]{$\nu_3$}
\Text(23,50)[l]{$\nu_5$}
\end{picture}
\\
\end{eqnarray*}
\caption{\label{fig1} 
\it Factorization in terms of diagrams:
The two-loop two-point function on the l.h.s. is equal to the insertion
of a one-loop three-point function into a one-loop two-point function.
The insertion occurs at the shaded vertex.}
\end{center}
\end{figure}
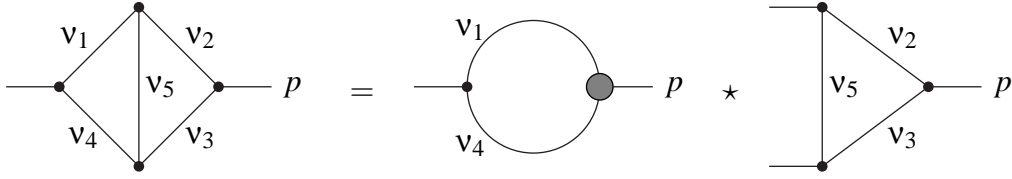
This is pictorially shown in Fig. (\ref{fig1}).
At the level of Feynman diagrams the product structure is given as 
the insertion of a Feynman diagram into another Feynman diagram.
It is a well known fact that convolution products can be turned into ordinary
products by applying a suitable integral transformation.
In the case at hand, eq. (\ref{convolutionprod})
factorizes by performing two inverse Mellin transformations
in $\nu_1$ and $\nu_4$.
If one sets
\bq
I^{(1,2)}(m-\eps,x,y)
 & = & 
 \frac{1}{(2 \pi i)^2}
 \int\limits_{\gamma_3-i \infty}^{\gamma_3+i \infty} d\nu_1 \int\limits_{\gamma_4-i \infty}^{\gamma_4+i \infty} d\nu_4 \;
 x^{-\nu_1} y^{-\nu_4} \;
 \hat{I}^{(1,2)}(m-\eps,\nu_1,\nu_4),
 \\
I^{(2,5)}(m-\eps,\nu_2,\nu_3,\nu_5;x,y)
 & = &
 \frac{1}{(2 \pi i)^2}
 \int\limits_{\gamma_3-i \infty}^{\gamma_3+i \infty} d\nu_1 \int\limits_{\gamma_4-i \infty}^{\gamma_4+i \infty} d\nu_4 \;
 x^{-\nu_1} y^{-\nu_4} \;
 \hat{I}^{(2,5)}(m-\eps,\nu_1,\nu_2,\nu_3,\nu_4,\nu_5), \nonumber 
\eq
one obtains the double inverse Mellin transform of the two-loop integral $\hat{I}^{(2,5)}$
as the product of the one-loop integral $I^{(1,3)}$ with the double inverse Mellin transform
of the one-loop integral $\hat{I}^{(1,2)}$:
\bq
\label{prodtrafo}
I^{(2,5)}(m-\eps,\nu_2,\nu_3,\nu_5;x,y)
 & = &
 I^{(1,2)}(m-\eps,x,y)
 \times
 I^{(1,3)}(m-\eps,\nu_2,\nu_3,\nu_5; x, y).
\eq
Eq. (\ref{convolutionprod}) or eq. (\ref{prodtrafo}) is the advertised factorization
of the two-loop integral into two one-loop integrals.
At the level of Feynman diagrams the product is similar to the insertion
operation defined by Kreimer within the context of renormalization
\cite{Kreimer:2002qy,Kreimer:2002fv,Kreimer:2003vp}.
In the context discussed here, insertions occur only at one specified place.
This implies that the product is associative.
Note that in general an insertion product, which allows insertions at
several places, is not associative.
The factorization property, e.g. that non-primitive graphs can be written
as convolution products of primitive graphs, generalizes to higher loops.
For the three-loop two-point functions there are three basic topologies,
usually named the ladder (``LA'') topology, the Benz (``BE'') topology
and the non-planar (``NO'') topology.
They are shown on the l.h.s of fig. (\ref{fig2}).
\begin{figure}
\begin{center}
\begin{eqnarray*}
\begin{picture}(80,35)(0,25)
\Vertex(10,30){2}
\Vertex(30,10){2}
\Vertex(30,50){2}
\Vertex(50,10){2}
\Vertex(50,50){2}
\Vertex(70,30){2}
\CArc(30,30)(20,90,270)
\CArc(50,30)(20,-90,90)
\Line(30,10)(50,10)
\Line(30,50)(50,50)
\Line(30,10)(30,50)
\Line(50,10)(50,50)
\Line(0,30)(10,30)
\Line(70,30)(80,30)
\Text(13,44)[br]{\small$\nu_1$}
\Text(13,17)[tr]{\small$\nu_6$}
\Text(67,44)[bl]{\small$\nu_3$}
\Text(67,17)[tl]{\small$\nu_4$}
\Text(40,52)[b]{\small$\nu_2$}
\Text(40,8)[t]{\small$\nu_5$}
\Text(28,30)[r]{\small$\nu_7$}
\Text(48,30)[r]{\small$\nu_8$}
\end{picture}
& = & 
\begin{picture}(60,35)(0,25)
\Vertex(10,30){2}
\Line(0,30)(10,30)
\Line(40,30)(50,30)
\CArc(25,30)(15,0,360)
\Text(15,44)[br]{\small$\nu_1$}
\Text(15,17)[tr]{\small$\nu_6$}
\GCirc(40,30){4}{0.5}
\end{picture}
\star \;\;\;
\begin{picture}(60,35)(0,25)
\Vertex(10,15){2}
\Vertex(10,45){2}
\Line(0,15)(10,15)
\Line(0,45)(10,45)
\Line(40,30)(50,30)
\Line(10,15)(10,45)
\Line(10,45)(40,30)
\Line(40,30)(10,15)
\Text(28,39)[bl]{\small$\nu_2$}
\Text(28,19)[tl]{\small$\nu_5$}
\Text(13,30)[l]{\small$\nu_7$}
\GCirc(40,30){4}{0.5}
\end{picture}
\star \;\;\;
\begin{picture}(60,35)(0,25)
\Vertex(10,15){2}
\Vertex(10,45){2}
\Vertex(40,30){2}
\Line(0,15)(10,15)
\Line(0,45)(10,45)
\Line(40,30)(50,30)
\Line(10,15)(10,45)
\Line(10,45)(40,30)
\Line(40,30)(10,15)
\Text(28,39)[bl]{\small$\nu_3$}
\Text(28,19)[tl]{\small$\nu_4$}
\Text(13,30)[l]{\small$\nu_8$}
\end{picture}
\nonumber \\
& & \nonumber \\
\begin{picture}(80,60)(0,25)
\Vertex(10,30){2}
\Vertex(40,10){2}
\Vertex(30,50){2}
\Vertex(40,30){2}
\Vertex(50,50){2}
\Vertex(70,30){2}
\CArc(30,30)(20,90,270)
\CArc(50,30)(20,-90,90)
\Line(30,10)(50,10)
\Line(30,50)(50,50)
\Line(0,30)(10,30)
\Line(70,30)(80,30)
\Line(40,10)(40,30)
\Line(40,30)(30,50)
\Line(40,30)(50,50)
\Text(13,44)[br]{\small$\nu_1$}
\Text(13,17)[tr]{\small$\nu_5$}
\Text(67,44)[bl]{\small$\nu_3$}
\Text(67,17)[tl]{\small$\nu_4$}
\Text(40,52)[b]{\small$\nu_2$}
\Text(42,20)[l]{\small$\nu_8$}
\Text(33,38)[r]{\small$\nu_6$}
\Text(48,38)[l]{\small$\nu_7$}
\end{picture}
& = & 
\begin{picture}(60,60)(0,25)
\Vertex(10,30){2}
\Line(0,30)(10,30)
\Line(40,30)(50,30)
\CArc(25,30)(15,0,360)
\Text(15,44)[br]{\small$\nu_1$}
\Text(15,17)[tr]{\small$\nu_5$}
\GCirc(40,30){4}{0.5}
\end{picture}
\star \;\;\;
\begin{picture}(60,60)(0,25)
\Vertex(10,15){2}
\Vertex(40,30){2}
\Line(0,15)(10,15)
\Line(0,45)(10,45)
\Line(40,30)(50,30)
\Line(10,15)(10,45)
\Line(10,45)(40,30)
\Line(40,30)(10,15)
\Text(28,39)[bl]{\small$\nu_3$}
\Text(28,19)[tl]{\small$\nu_4$}
\Text(13,30)[l]{\small$\nu_8$}
\GCirc(10,45){4}{0.5}
\end{picture}
\star \;\;\;
\begin{picture}(60,60)(0,25)
\Vertex(10,15){2}
\Vertex(10,45){2}
\Vertex(40,15){2}
\Line(0,45)(10,45)
\Line(10,5)(10,45)
\Line(10,45)(50,5)
\Line(40,15)(10,15)
\Line(10,45)(10,5)
\Text(28,29)[bl]{\small$\nu_2$}
\Text(30,11)[t]{\small$\nu_7$}
\Text(13,25)[l]{\small$\nu_6$}
\end{picture}
\nonumber \\
& & \nonumber \\
\begin{picture}(80,60)(0,25)
\Vertex(10,30){2}
\Vertex(30,10){2}
\Vertex(30,50){2}
\Vertex(50,10){2}
\Vertex(50,50){2}
\Vertex(70,30){2}
\CArc(30,30)(20,90,270)
\CArc(50,30)(20,-90,90)
\Line(30,10)(50,10)
\Line(30,50)(50,50)
\Line(0,30)(10,30)
\Line(70,30)(80,30)
\Line(30,10)(39,28)
\Line(41,32)(50,50)
\Line(50,10)(30,50)
\Text(13,44)[br]{\small$\nu_1$}
\Text(13,17)[tr]{\small$\nu_6$}
\Text(67,44)[bl]{\small$\nu_3$}
\Text(67,17)[tl]{\small$\nu_4$}
\Text(40,52)[b]{\small$\nu_2$}
\Text(40,8)[t]{\small$\nu_5$}
\Text(33,38)[r]{\small$\nu_7$}
\Text(48,38)[l]{\small$\nu_8$}
\end{picture}
& = & 
\begin{picture}(60,60)(0,25)
\Vertex(10,30){2}
\Line(0,30)(10,30)
\Line(40,30)(50,30)
\CArc(25,30)(15,0,360)
\Text(15,44)[br]{\small$\nu_1$}
\Text(15,17)[tr]{\small$\nu_6$}
\GCirc(40,30){4}{0.5}
\end{picture}
\star \;\;\;
\begin{picture}(60,60)(0,25)
\Line(50,30)(60,30)
\Vertex(50,30){2}
\Line(0,5)(50,30)
\Line(0,55)(50,30)
\Vertex(10,10){2}
\Vertex(10,50){2}
\Vertex(26,18){2}
\Vertex(26,42){2}
\Line(10,50)(26,18)
\Line(10,10)(19,28)
\Line(21,32)(26,42)
\Text(37,38)[bl]{\small$\nu_3$}
\Text(37,20)[tl]{\small$\nu_4$}
\Text(18,47)[bl]{\small$\nu_2$}
\Text(18,11)[tl]{\small$\nu_5$}
\Text(14,33)[r]{\small$\nu_7$}
\Text(24,32)[l]{\small$\nu_8$}
\end{picture}
\nonumber \\
\end{eqnarray*}
\caption{\label{fig2} 
\it Factorization at three loops:
The first two topologies factorize into one-loop diagrams, whereas the last
topology factorizes into a one-loop diagram and a two-loop diagram.
A graph on the right side of the product operator $\star$ is inserted into 
the shaded vertex of the graph to the left of the product operator.
}
\end{center}
\end{figure}
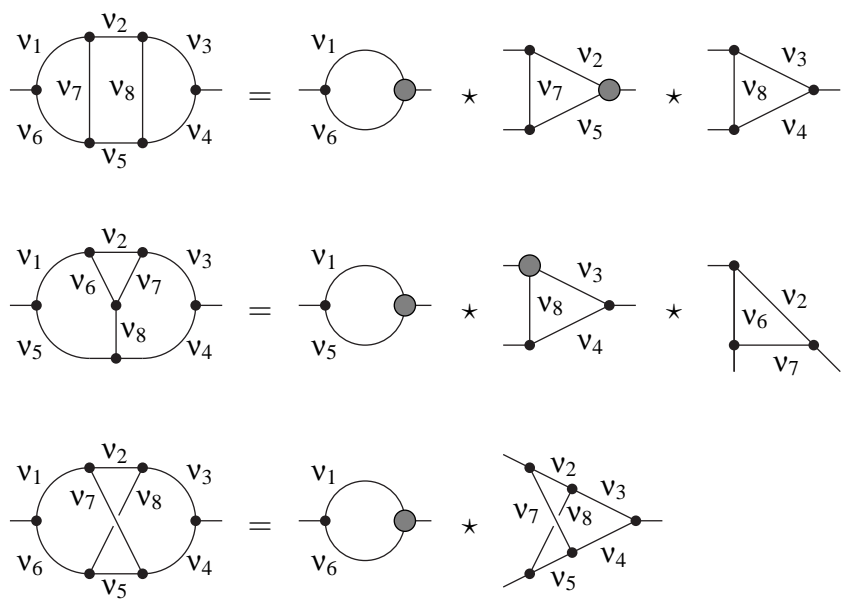
The ladder and Benz topologies are given as convolution products of
three one-loop graphs:
\pagebreak
\bq
\label{ladder}
\lefteqn{
\hat{I}^{(3,8)}_{LA}(m-\eps,\nu_1,\nu_2,\nu_3,\nu_4,\nu_5,\nu_6,\nu_7,\nu_8)
 } \nonumber \nopagebreak[4] \\
\nopagebreak[4] 
 & = &
 c_\Gamma^{-3}
 \left( -p^2 \right)^{\nu_{12345678}-3m+3\eps} 
 \int \frac{d^Dk_1}{i \pi^{D/2}}
 \int \frac{d^Dk_2}{i \pi^{D/2}}
 \int \frac{d^Dk_3}{i \pi^{D/2}}
 \prod\limits_{j=1}^8
  \frac{1}{ \left(-k_j^2\right)^{\nu_j}
          }
 \nonumber \\
 & = &
 \frac{1}{(2 \pi i)^4}
 \int\limits_{\gamma_1-i \infty}^{\gamma_1+i \infty} d\sigma_1 
 \int\limits_{\gamma_2-i \infty}^{\gamma_2+i \infty} d\tau_1 \;
 \int\limits_{\gamma_3-i \infty}^{\gamma_3+i \infty} d\sigma_2 
 \int\limits_{\gamma_4-i \infty}^{\gamma_4+i \infty} d\tau_2 \;
  \hat{I}^{(1,2)}(m-\eps,\nu_1-\tau_1,\nu_6-\sigma_1)
\nonumber \\
 & &
 \times \;
  \hat{I}^{(1,3)}(m-\eps,\nu_2-\tau_2,\nu_5-\sigma_2,\nu_7; \tau_1, \sigma_1)
  \hat{I}^{(1,3)}(m-\eps,\nu_3,\nu_4,\nu_8; \tau_2, \sigma_2).
\eq
The momenta are defined for the ladder topology by 
$k_4=k_3-p$, $k_5=k_2-p$, $k_6=k_1-p$, $k_7=k_2-k_1$ and $k_8=k_3-k_2$.
For the Benz topology we have
\bq
\label{benz}
\lefteqn{
\hat{I}^{(3,8)}_{BE}(m-\eps,\nu_1,\nu_2,\nu_3,\nu_4,\nu_5,\nu_6,\nu_7,\nu_8)
 } \nonumber \\
 & = &
 c_\Gamma^{-3}
 \left( -p^2 \right)^{\nu_{12345678}-3m+3\eps} 
 \int \frac{d^Dk_1}{i \pi^{D/2}}
 \int \frac{d^Dk_2}{i \pi^{D/2}}
 \int \frac{d^Dk_3}{i \pi^{D/2}}
 \prod\limits_{j=1}^8
  \frac{1}{ \left(-k_j^2\right)^{\nu_j}
          }
 \nonumber \\
 & = &
 \frac{1}{(2 \pi i)^4}
 \int\limits_{\gamma_1-i \infty}^{\gamma_1+i \infty} d\sigma_1 
 \int\limits_{\gamma_2-i \infty}^{\gamma_2+i \infty} d\tau_1 \;
 \int\limits_{\gamma_3-i \infty}^{\gamma_3+i \infty} d\sigma_2 
 \int\limits_{\gamma_4-i \infty}^{\gamma_4+i \infty} d\tau_2 \;
  \hat{I}^{(1,2)}(m-\eps,\nu_1-\tau_1-\tau_2,\nu_5-\sigma_1)
 \\
 & &
 \times \;
  \hat{I}^{(1,3)}(m-\eps,\nu_{2367}-m+\eps+\sigma_2+\tau_2,\nu_4,\nu_8-\sigma_2; \tau_1, \sigma_1)
  \hat{I}^{(1,3)}(m-\eps,\nu_2,\nu_7,\nu_6; \tau_2, \sigma_2).
 \nonumber 
\eq
The momenta are defined for the Benz topology by 
$k_4=k_3-p$, $k_5=k_1-p$, $k_6=k_2-k_1$, $k_7=k_3-k_2$ and $k_8=k_3-k_1$.
The non-planar topology factorizes only into a one-loop two-point function
and the crossed two-loop three-point function:
\bq
\label{nonplanar}
\lefteqn{
\hat{I}^{(3,8)}_{NO}(m-\eps,\nu_1,\nu_2,\nu_3,\nu_4,\nu_5,\nu_6,\nu_7,\nu_8)
 } \nonumber \\
 & = &
 c_\Gamma^{-3}
 \left( -p^2 \right)^{\nu_{12345678}-3m+3\eps} 
 \int \frac{d^Dk_1}{i \pi^{D/2}}
 \int \frac{d^Dk_2}{i \pi^{D/2}}
 \int \frac{d^Dk_3}{i \pi^{D/2}}
 \prod\limits_{j=1}^8
  \frac{1}{ \left(-k_j^2\right)^{\nu_j}
          }
 \\
 & = &
 \frac{1}{(2 \pi i)^2}
 \int\limits_{\gamma_1-i \infty}^{\gamma_1+i \infty} d\sigma 
 \int\limits_{\gamma_2-i \infty}^{\gamma_2+i \infty} d\tau \;
  \hat{I}^{(1,2)}(m-\eps,\nu_1-\tau,\nu_6-\sigma)
  \hat{I}^{(2,6)}_X(m-\eps,\nu_2,\nu_3,\nu_4,\nu_5,\nu_7,\nu_8;\tau,\sigma).
 \nonumber 
\eq
Here, $\hat{I}^{(2,6)}_X$ is the double Mellin transform 
\bq
\lefteqn{
\hat{I}^{(2,6)}_X(m-\eps,\nu_2,\nu_3,\nu_4,\nu_5,\nu_7,\nu_8;\tau,\sigma)
 = } \nonumber \\
 & &
 \int\limits_0^\infty dx 
 \int\limits_0^\infty dy \;
 x^{\tau - 1} y^{\sigma -1} \;
I^{(2,6)}_X(m-\eps,\nu_2,\nu_3,\nu_4,\nu_5,\nu_7,\nu_8;x,y)
\eq
of the two-loop crossed vertex function
\bq
\label{crossedvertex}
\lefteqn{
I^{(2,6)}_X(m-\eps,\nu_2,\nu_3,\nu_4,\nu_5,\nu_7,\nu_8;x,y)
 = } \\
 & &
 c_\Gamma^{-2}
 \left( -p^2 \right)^{\nu_{234578}-2m+2\eps} 
 \int \frac{d^Dk_2}{i \pi^{D/2}}
 \int \frac{d^Dk_3}{i \pi^{D/2}}
  \frac{1}{ \left(-k_2^2\right)^{\nu_2}
            \left(-k_3^2\right)^{\nu_3}
            \left(-k_4^2\right)^{\nu_4}
            \left(-k_5^2\right)^{\nu_5}
            \left(-k_7^2\right)^{\nu_7}
            \left(-k_8^2\right)^{\nu_8}
          }. \nonumber 
\eq
The momenta for eq. (\ref{nonplanar}) and eq. (\ref{crossedvertex})
are defined by
$k_4=k_3-p$, $k_5=k_1-k_2+k_3-p$, $k_6=k_1-p$, $k_7=k_2-k_1$ and
$k_8=k_3-k_2$.
The variables $x$ and $y$ are defined by
$x=(-p^2)/(-k_1^2)$ and $y=(-p^2)/(-k_6^2)$.
For all topologies the factorization is shown pictorially in fig. (\ref{fig2}).

\section{Evaluation of the two-loop integral}
\label{sect:eval}

Eq. (\ref{convolutionprod}) 
is the starting point for the further evaluation of the
two-loop integral:
\bq
\label{tobesolved}
\lefteqn{
\hat{I}^{(2,5)}
 = 
 c \frac{1}{(2 \pi i)^2}
 \int\limits_{\gamma_1-i \infty}^{\gamma_1+i \infty} d\sigma 
 \int\limits_{\gamma_2-i \infty}^{\gamma_2+i \infty} d\tau 
 \;
 \frac{\Gamma(-\sigma) \Gamma(-\sigma+m-\eps-\nu_{35}) \Gamma(\sigma+m-\eps-\nu_4)}{\Gamma(-\sigma+\nu_4)}
 } \nonumber \\
 & & \times
 \frac{\Gamma(-\tau) \Gamma(-\tau+m-\eps-\nu_{25}) \Gamma(\tau+m-\eps-\nu_1)}{\Gamma(-\tau+\nu_1)}
 \nonumber \\
 & & \times
 \frac{\Gamma(-\sigma-\tau-m+\eps+\nu_{14}) \Gamma(\sigma+\tau-m+\eps+\nu_{235}) \Gamma(\sigma+\tau+\nu_5)}{\Gamma(\sigma+\tau+2m-2\eps-\nu_{14})},
\eq
with
\bq
c & = &  
 \frac{c_\Gamma^{-2}}{\Gamma(\nu_2)\Gamma(\nu_3)\Gamma(\nu_5)\Gamma(2m-2\eps-\nu_{235})}.
\eq
The strategy for the evaluation is as follows: We first close the contours and evaluate the integrals
with the help of the residuum theorem.
This technique 
has a long history, a recent example is the calculation of the triple box \cite{Smirnov:2003vi}. 
The semi-circles at infinity needed to close the contours give a vanishing contribution
provided that
\bq
\label{ineq}
\nu_1 + \nu_{125} -2m + 2\eps & < & 1,
\nonumber \\
\nu_4 + \nu_{345} -2m + 2\eps & < & 1,
\nonumber \\
-1 & < & \left( \nu_1 + \nu_{125} -2m + 2\eps \right)
        +\left( \nu_4 + \nu_{345} -2m + 2\eps \right).
\eq
From the residues we immediately obtain nested sums. All these sums can be brought to a standard form
and then expanded into a Laurent series in $\eps$ with the help of the algorithms described in
\cite{Moch:2001zr}. We use the program ``nestedsums'' \cite{Weinzierl:2002hv}
and the ``GiNaC''-library \cite{Bauer:2000cp} for this purpose.
The algorithms A and B of \cite{Moch:2001zr} are generalizations of algorithms for
the manipulations of harmonic sums and harmonic polylogarithms, described in 
\cite{Vermaseren:1998uu,Remiddi:1999ew,Blumlein:1998if}.

Closing the contour of the $\sigma$-integration
to the right, one picks up the residues of $\Gamma(-\sigma)$,
$\Gamma(-\sigma+m-\eps-\nu_{35})$ and $\Gamma(-\sigma-\tau-m+\eps+\nu_{14})$.
All other $\sigma$-dependent Gamma functions in the numerator have poles to the left
of the contour and therefore do not contribute.
The basic formula for the residuum of Euler's Gamma-function reads:
\bq
\mbox{res} \; \left( \Gamma(-x+a), x=a+n \right) & = & -\frac{(-1)^n}{n!} 
\eq
The two-loop integral can therefore be written as a sum of three terms,
\bq
\hat{I}^{(2,5)}
 & = &
 T_{(1)} + T_{(2)} + T_{(3)},
\eq
where each term is obtained by taking the residues of one Gamma function from
the set $\Gamma(-\sigma)$,
$\Gamma(-\sigma+m-\eps-\nu_{35})$ and $\Gamma(-\sigma-\tau-m+\eps+\nu_{14})$.
Explicitly,
\bq
T_{(k)} & = &
 c 
 \sum\limits_{n=0}^\infty \frac{(-1)^n}{n!}
 \frac{1}{2 \pi i} 
 \int\limits_{\gamma_2-i \infty}^{\gamma_2+i \infty} d\tau 
 \;
 \frac{\Gamma(-\tau) \Gamma(-\tau+m-\eps-\nu_{25}) \Gamma(\tau+m-\eps-\nu_1)}{\Gamma(-\tau+\nu_1)}
 \; H_{(k)},
\eq
with
\bq
H_{(1)} & = &
 \frac{\Gamma(-n+m-\eps-\nu_{35}) \Gamma(n+m-\eps-\nu_4)}{ \Gamma(-n+\nu_4)}
 \nonumber \\
 & &
 \times
 \frac{ \Gamma(-\tau-n-m+\eps+\nu_{14}) \Gamma(\tau+n-m+\eps+\nu_{235}) \Gamma(\tau+n+\nu_5)}{\Gamma(\tau+n+2 m- 2 \eps - \nu_{14})}, 
 \nonumber \\
H_{(2)} & = &
 \frac{\Gamma(-n-m+\eps+\nu_{35}) \Gamma(n+2m-2\eps-\nu_{345})}{\Gamma(-n-m+\eps+\nu_{345})}
 \nonumber \\
 & &
 \times
 \frac{\Gamma(-\tau-n-2 m + 2 \eps +\nu_{1345}) \Gamma(\tau+n+\nu_2) \Gamma(\tau+n+m-\eps-\nu_3)}{\Gamma(\tau+n+3 m - 3 \eps -\nu_{1345})},
 \nonumber \\
H_{(3)} & = & 
 \frac{ \Gamma(n-2 m + 2 \eps + \nu_{12345}) \Gamma(n-m+\eps+\nu_{145})}{\Gamma(n+m-\eps)}
 \nonumber \\
 & & 
 \times
 \frac{\Gamma(\tau-n+m-\eps-\nu_{14}) \Gamma(\tau-n+2 m -2 \eps -\nu_{1345}) \Gamma(-\tau+n+\nu_1)}{\Gamma(\tau-n+m-\eps-\nu_1)}.
\eq 
The procedure is then repeated by closing the $\tau$-integration contour
to the right.
For example, for the term $T_{(1)}$ we have to evaluate the residues of
$\Gamma(-\tau)$, $\Gamma(-\tau+m-\eps-\nu_{25})$ and $\Gamma(-\tau-n-m+\eps+\nu_{14})$.
The residues of $\Gamma(\tau+m-\eps-\nu_1)$, $\Gamma(\tau+n-m+\eps+\nu_{235})$ or
$\Gamma(\tau+n+\nu_5)$ are always to the left of the contour.
In summary, the two-loop integral can be written as the sum of the following terms:
\bq
\hat{I}^{(2,5)}
 & = &
   T_{(1,1)} + T_{(1,2)} + T_{(1,3)}
 + T_{(2,1)} + T_{(2,2)} + T_{(2,3)}
 + T_{(3,1)} + T_{(3,2)} + T_{(3,3)} + T_{(3,4)} + T_{(3,5)}.
 \nonumber \\
\eq
Table (\ref{tabresidues}) shows the 
correspondence between each term $T_{(k,l)}$ and the Gamma functions, from
which the residues are taken.
\begin{table}
\begin{center}
\begin{tabular}{|c|ll|} \hline
 & $\sigma$ & $\tau$ \\
\hline 
 $T_{(1,1)}$ & $\Gamma(-\sigma)$ & $\Gamma(-\tau)$ \\
 $T_{(1,2)}$ & $\Gamma(-\sigma)$ & $\Gamma(-\tau+m-\eps-\nu_{25})$ \\
 $T_{(1,3)}$ & $\Gamma(-\sigma)$ & $\Gamma(-\tau-n-m+\eps+\nu_{14})$ \\
 & & \\
 $T_{(2,1)}$ & $\Gamma(-\sigma+m-\eps-\nu_{35})$ & $\Gamma(-\tau)$ \\
 $T_{(2,2)}$ & $\Gamma(-\sigma+m-\eps-\nu_{35})$ & $\Gamma(-\tau+m-\eps-\nu_{25})$ \\
 $T_{(2,3)}$ & $\Gamma(-\sigma+m-\eps-\nu_{35})$ & $\Gamma(-\tau-n-2 m+2 \eps+\nu_{1345})$ \\
 & & \\
 $T_{(3,1)}$ & $\Gamma(-\sigma-\tau-m+\eps+\nu_{14})$ & $\Gamma(-\tau)$ \\
 $T_{(3,2)}$ & $\Gamma(-\sigma-\tau-m+\eps+\nu_{14})$ & $\Gamma(-\tau+m-\eps-\nu_{25})$ \\
 $T_{(3,3)}$ & $\Gamma(-\sigma-\tau-m+\eps+\nu_{14})$ & $\Gamma(-\tau+n+\nu_1)$ \\
 $T_{(3,4)}$ & $\Gamma(-\sigma-\tau-m+\eps+\nu_{14})$ & $\Gamma(\tau-n+m-\eps-\nu_{14})$ \\
 $T_{(3,5)}$ & $\Gamma(-\sigma-\tau-m+\eps+\nu_{14})$ & $\Gamma(\tau-n+2 m - 2 \eps - \nu_{1345})$ \\
\hline
\end{tabular}
\end{center}
\caption{\label{tabresidues}
\it Correspondence between the terms $T_{(k,l)}$ and the Gamma functions, from
which the residues are taken.}
\end{table}
The term $T_{(3,3)}$ corresponding to the residues in $\Gamma(-\tau+n+\nu_1)$ gives
no contribution.
This Gamma function is always accompanied by $1/\Gamma(-\tau+\nu_1)$.
Since
\bq
\frac{\Gamma(-\tau+n+\nu_1)}{\Gamma(-\tau+\nu_1)} & = & (-\tau+\nu_1+n-1) ... (-\tau+\nu_1)
\eq
is free of poles we have
\bq
T_{(3,3)} & = & 0.
\eq
Furthermore, we can show that the terms $T_{(1,3)}$ and $T_{(3,4)}$ as well as
the terms $T_{(2,3)}$ and $T_{(3,5)}$ 
cancel each other:
\bq
T_{(1,3)} + T_{(3,4)} = 0,
& &
T_{(2,3)} + T_{(3,5)} = 0.
\eq
To present the results after all residues have been taken, we introduce two functions $G_\pm$ with
ten arguments each:
\bq
\lefteqn{
G_\pm(a_1,a_2,a_3,a_4;b_1,b_2,b_3;c_1,c_2,c_3) = }
\nonumber \\
& & 
 \sum\limits_{n=0}^\infty 
 \sum\limits_{j=0}^\infty
  \frac{(-1)^{n+j}}{n! j!}
  \frac{\Gamma(\mp n - j-a_1) \Gamma( \pm n + j+a_2) \Gamma( \pm n + j+a_3)}{\Gamma( \pm n + j+a_4)}
 \nonumber \\
 & &
  \times
  \frac{\Gamma(\mp n \mp b_1) \Gamma(n+b_2)}{\Gamma(\mp n \mp b_3)}
  \frac{\Gamma(-j-c_1) \Gamma(j+c_2)}{\Gamma(-j-c_3)},
\eq
together with two operators ${\cal L}_{d}$ and ${\cal R}_{d}$ acting on the arguments as follows:
\bq
\lefteqn{
{\cal L}_{d} G_\pm(a_1,a_2,a_3,a_4;b_1,b_2,b_3;c_1,c_2,c_3)
 = }
 \nonumber \\
 & &
 G_\pm(a_1+d,a_2+d,a_3+d,a_4+d;b_1+2 d,b_2+d,b_3+d;c_1,c_2,c_3),
 \nonumber \\
\lefteqn{
{\cal R}_{d} G_\pm(a_1,a_2,a_3,a_4;b_1,b_2,b_3;c_1,c_2,c_3)
 = }
 \nonumber \\
 & &
 G_\pm(a_1+d,a_2+d,a_3+d,a_4+d;b_1,b_2,b_3;c_1+2 d,c_2+d ,c_3+d).
\eq
Then
\bq
\label{cressum}
\lefteqn{
\hat{I}^{(2,5)}
 = c
 \left( {\bf 1} + {\cal L}_{m-\eps-\nu_{35}} + {\cal R}_{m-\eps-\nu_{25}} 
                + {\cal L}_{m-\eps-\nu_{35}} {\cal R}_{m-\eps-\nu_{25}} 
 \right)
}
\nonumber \\
 & &
 G_+\left(m-\eps-\nu_{14}, -m+\eps+\nu_{235}, \nu_5, 2m-2\eps-\nu_{14};
          -m+\eps+\nu_{35}, m-\eps-\nu_4, -\nu_4;
 \right. \nonumber \\
 & & \left. \;\;\;\;\;\;\;\;
          -m+\eps+\nu_{25}, m-\eps-\nu_1, -\nu_1
    \right)
 \nonumber \\
 & &
 + c
 \left( {\bf 1} + {\cal R}_{m-\eps-\nu_{25}} 
 \right)
 G_-\left( -\nu_1, m-\eps-\nu_{14}, 2m-2\eps-\nu_{1345}, m-\eps-\nu_1;
 \right. \nonumber \\
 & & \left. \;\;\;\;\;\;\;\;
          -2m+2\eps+\nu_{12345}, -m+\eps+\nu_{145}, m-\eps;
          -m+\eps+\nu_{25}, m-\eps-\nu_1, -\nu_1
    \right).
\eq
Here, ${\bf 1}$ denotes the identity operator with a trivial action on the arguments of the functions $G_\pm$.
To proceed further, we now show how to transform the functions $G_\pm$ to a standard
form, such that they can be expanded in $\eps$ with algorithms of
\cite{Moch:2001zr,Weinzierl:2002hv}.
Since the shift operators ${\cal L}$ and ${\cal R}$ modify only the arguments
but not the structure of the functions $G_\pm$,
it is sufficient to discuss one example for each function.
For the function $G_+$ we discuss as an example the first term (without any application of the shift
operators ${\cal L}$ or ${\cal R}$) of eq. (\ref{cressum}), which corresponds to the term $T_{(1,1)}$.
\bq
\label{T11}
\lefteqn{
T_{(1,1)}
 = c \;
 G_+(m-\eps-\nu_{14}, -m+\eps+\nu_{235}, \nu_5, 2m-2\eps-\nu_{14};
          -m+\eps+\nu_{35}, m-\eps-\nu_4, -\nu_4;
 } \nonumber \\
 & & \;\;\;\;\;\;\;\;
          -m+\eps+\nu_{25}, m-\eps-\nu_1, -\nu_1
    )
 \nonumber \\
 & = &
 c
 \sum\limits_{n=0}^\infty \sum\limits_{j=0}^\infty
 \frac{(-1)^{(n+j)}}{n! j!}
 \frac{\Gamma(-n-j-m+\eps+\nu_{14}) \Gamma(n+j-m+\eps+\nu_{235}) \Gamma(n+j+\nu_5)}{\Gamma(n+j+2m-2\eps-\nu_{14})}
 \nonumber \\ 
 & & \times 
 \frac{\Gamma(-n+m-\eps-\nu_{35}) \Gamma(n+m-\eps-\nu_4)}{\Gamma(-n+\nu_4)}
 \frac{\Gamma(-j+m-\eps-\nu_{25}) \Gamma(j+m-\eps-\nu_1)}{\Gamma(-j+\nu_1)}.
\eq
If $\nu_1$ and $\nu_4$ are not positive integers, one may use the reflection formula
\bq
\Gamma(-n+x) & = & \Gamma(x) \Gamma(1-x) \frac{(-1)^n}{\Gamma(n+1-x)}
\eq
to flip the Gamma functions, where a summation index occurs with a negative sign.
One obtains in that case
\bq
\label{T11noint}
T_{(1,1)}
 & = & 
 c \; \Gamma(-m+\eps+\nu_{14}) \Gamma(1+m-\eps-\nu_{14})
 \nonumber \\
 & &
 \times
 \frac{\Gamma(m-\eps-\nu_{35}) \Gamma(1-m+\eps+\nu_{35}) 
       \Gamma(m-\eps-\nu_{25}) \Gamma(1-m+\eps+\nu_{25})
     }{\Gamma(\nu_1)\Gamma(1-\nu_1) \Gamma(\nu_4)\Gamma(1-\nu_4)}
 \nonumber \\
 & &
 \times 
 \sum\limits_{n=0}^\infty \sum\limits_{j=0}^\infty
 \frac{\Gamma(n+m-\eps-\nu_4) \Gamma(n+1-\nu_4)}{\Gamma(n+1) \Gamma(n+1-m+\eps+\nu_{35})}
 \frac{\Gamma(j+m-\eps-\nu_1) \Gamma(j+1-\nu_1)}{\Gamma(j+1) \Gamma(j+1-m+\eps+\nu_{25})}
 \nonumber \\
 & &
 \times
 \frac{\Gamma(n+j-m+\eps+\nu_{235}) \Gamma(n+j+\nu_5)}{\Gamma(n+j+2m-2\eps-\nu_{14}) \Gamma(n+j+1+m-\eps-\nu_{14})}.
\eq
This is a double infinite sum with unit arguments 
and can be evaluated with algorithm B of \cite{Moch:2001zr,Weinzierl:2002hv}.
The sum in eq.(\ref{T11noint}) is a generalization of the first Appell function.
The Laurent expansion to any fixed order will contain only rational numbers
and multiple zeta values.
If $\nu_1$ or $\nu_4$ are positive integers, the infinite sums over $j$ or $n$ terminate.
For example, if $\nu_1$ is a positive integer, while $\nu_4$ is not, one obtains
\bq
\label{T11oneint}
\lefteqn{
T_{(1,1)}
 =  
 c
 \;
 \frac{\Gamma(m-\eps-\nu_{35}) \Gamma(1-m+\eps+\nu_{35}) \Gamma(-m+\eps+\nu_{14}) \Gamma(1+m-\eps-\nu_{14})}{\Gamma(\nu_4) \Gamma(1-\nu_4)}
} \nonumber \\
 & &
 \times
 \sum\limits_{j=0}^{\nu_1-1}
 \frac{\Gamma(-j+m-\eps-\nu_{25}) \Gamma(j+m-\eps-\nu_1)}{\Gamma(j+1) \Gamma(-j+\nu_1)}
 \sum\limits_{n=0}^\infty 
 \frac{\Gamma(n+m-\eps-\nu_4) \Gamma(n+1-\nu_4)}{\Gamma(n+1) \Gamma(n+1-m+\eps+\nu_{35})}
 \nonumber \\
 & & 
 \times
 \frac{\Gamma(n+j-m+\eps+\nu_{235}) \Gamma(n+j+\nu_5)}{\Gamma(n+j+2m-2\eps-\nu_{14}) \Gamma(n+j+1+m-\eps-\nu_{14})}.
\eq
This is a finite sum of single infinite sums, more precisely, it is a sum
of $\nu_1$ terms, each containing a ${}_4F_3$ hypergeometric function with unit argument.
These are evaluated with algorithm A of \cite{Moch:2001zr,Weinzierl:2002hv}.
Again,
the Laurent expansion to any fixed order will contain only rational numbers
and multiple zeta values.
The case where $\nu_4$ is a positive integer while $\nu_1$ is not, is completely analog.
If both $\nu_1$ and $\nu_4$ are positive integers, one obtains
\bq
\lefteqn{
T_{(1,1)}
 =  
 c
 \sum\limits_{n=0}^{\nu_4-1} \sum\limits_{j=0}^{\nu_1-1}
 \frac{(-1)^{(n+j)}}{n! j!}
 \frac{\Gamma(-n-j+\nu_{14}-m+\eps) \Gamma(n+j-m+\eps+\nu_{235}) \Gamma(n+j+\nu_5)}{\Gamma(n+j+2m-2\eps-\nu_{14})}
}
 \nonumber \\ 
 & & \times 
 \frac{\Gamma(-n+m-\eps-\nu_{35}) \Gamma(n+m-\eps-\nu_4)}{\Gamma(-n+\nu_4)}
 \frac{\Gamma(-j+m-\eps-\nu_{25}) \Gamma(j+m-\eps-\nu_1)}{\Gamma(-j+\nu_1)},
\eq
e.g. a finite sum of products of Gamma functions.
This is rather trivial and the Laurent expansion contains only
rational numbers
and zeta values.

We now turn to the function $G_-$. As an example we discuss
\bq
\label{T31}
\lefteqn{
T_{(3,1)} = 
c \;
 G_-\left( -\nu_1, m-\eps-\nu_{14}, 2m-2\eps-\nu_{1345}, m-\eps-\nu_1;
 \right. } \nonumber \\
 & & \left. \;\;\;\;\;\;\;\;
          -2m+2\eps+\nu_{12345}, -m+\eps+\nu_{145}, m-\eps;
          -m+\eps+\nu_{25}, m-\eps-\nu1, -\nu_1
    \right)
 \nonumber \\
& = &
 c
 \sum\limits_{n=0}^\infty
 \sum\limits_{j=0}^\infty
 \frac{(-1)^{n+j}}{n!j!}
 \frac{\Gamma(n-2 m + 2 \eps+\nu_{12345}) \Gamma(n-m+\eps+\nu_{145})}{\Gamma(n+m-\eps)}
 \nonumber \\
 & &
 \times
 \frac{\Gamma(-j+m-\eps-\nu_{25}) \Gamma(j+m-\eps-\nu_1)}{\Gamma(-j+\nu_1)}
 \nonumber \\
 & & 
 \times
 \frac{\Gamma(-n+j+m-\eps-\nu_{14}) \Gamma(-n+j+2 m -2 \eps-\nu_{1345}) \Gamma(n-j+\nu_1)}{\Gamma(-n+j+m-\eps-\nu_1)}.
\eq
If $\nu_1$ is not a positive integer, we split the summation region into the regions $j \le n$
and $n \le j$ and subtract the double counted diagonal $j=n$ as in
\bq
 \sum\limits_{n=0}^\infty
 \sum\limits_{j=0}^\infty F(n,j) & = &
  \sum\limits_{n=0}^\infty
 \sum\limits_{j=0}^n F(n,j)
+
 \sum\limits_{j=0}^\infty
 \sum\limits_{n=0}^j F(n,j)
-
 \sum\limits_{n=0}^\infty F(n,n).
\eq
We obtain
\bq
\lefteqn{
T_{(3,1)} =   c \; \Gamma(m-\eps-\nu_{14}) \Gamma(1-m+\eps+\nu_{14})
} \nonumber \\
 & & 
 \times
 \frac{ \Gamma(m-\eps-\nu_{25}) \Gamma(1-m+\eps+\nu_{25}) 
        \Gamma(2m-2\eps-\nu_{1345}) \Gamma(1-2m+2\eps+\nu_{1345}) }
      { \Gamma(\nu_1) \Gamma(1-\nu_1) \Gamma(m-\eps-\nu_1) \Gamma(1-m+\eps+\nu_1) }
 \nonumber \\
 & & 
 \times
 \sum\limits_{n=0}^\infty
 \sum\limits_{j=0}^\infty
 \frac{\Gamma(n+\nu_1) \Gamma(n+1-m+\eps+\nu_1)}
      {\Gamma(n+1-m+\eps+\nu_{14}) \Gamma(n+1-2m+2\eps+\nu_{1345})}
 \frac{\Gamma(j+m-\eps-\nu_1) \Gamma(j+1-\nu_1)}
      {\Gamma(j+1) \Gamma(j+1-m+\eps+\nu_{25})}
 \nonumber \\
 & &
 \times
 \frac{\Gamma(n+j-m+\eps+\nu_{145}) \Gamma(n+j-2m+2\eps+\nu_{12345})}
      {\Gamma(n+j+1) \Gamma(n+j+m-\eps)}
 \nonumber \\
 & & 
 + c \; \Gamma(m-\eps-\nu_{25}) \Gamma(1-m+\eps+\nu_{25})
 \sum\limits_{n=0}^\infty
 \sum\limits_{j=0}^\infty
 \frac{\Gamma(n+m-\eps-\nu_{14}) \Gamma(n+2m-2\eps-\nu_{1345})}
      {\Gamma(n+1-\nu_1) \Gamma(n+m-\eps-\nu_1)}
 \nonumber \\
 & & 
 \times
 \frac{\Gamma(j-m+\eps+\nu_{145}) \Gamma(j-2m+2\eps+\nu_{12345})}
      {\Gamma(j+1) \Gamma(j+m-\eps)}
 \frac{\Gamma(n+j+m-\eps-\nu_1) \Gamma(n+j+1-\nu_1)}
      {\Gamma(n+j+1) \Gamma(n+j+1-m+\eps+\nu_{25})}
 \nonumber \\
 & & 
 - c \; 
 \frac{\Gamma(m-\eps-\nu_{25}) \Gamma(1-m+\eps+\nu_{25}) \Gamma(m-\eps-\nu_{14})
       \Gamma(2m-2\eps-\nu_{1345})}
      {\Gamma(1-\nu_1) \Gamma(m-\eps-\nu_1)}
 \nonumber \\
 & &
 \times
 \sum\limits_{n=0}^\infty
 \frac{\Gamma(n+m-\eps-\nu_1) \Gamma(n-m+\eps+\nu_{145}) \Gamma(n-2m+2\eps+\nu_{12345})
       \Gamma(n+1-\nu_1)}
      {\Gamma(n+1) \Gamma(n+1) \Gamma(n+m-\eps) \Gamma(n+1-m+\eps+\nu_{25})} 
\eq
The sums are of the same type as discussed in eq. (\ref{T11noint}) and 
eq. (\ref{T11oneint}), therefore the Laurent expansion contains only rational
numbers and multiple zeta values.

If $\nu_1$ is a positive integer, we observe that in eq. (\ref{T31}) we have the building block 
\bq
\frac{\Gamma(n-j+\nu_1)}{\Gamma(-j+\nu_1)} & = &
 (\nu_1-j) (\nu_1-j+1) ... (\nu_1-j+n-1).
\eq
Therefore we obtain a non-vanishing contribution only if this
sequence does not contain zero. This is the case for
\bq
\mbox{Region I:} & & j < \nu_1,
\nonumber \\
\mbox{Region II:} & & j > n + \nu_1 -1.
\eq
Region I contains only a finite number of residues from $\Gamma(-\tau)$, whereas region II
contains an infinite sum.
We obtain 
\bq
\lefteqn{
T_{(3,1)} 
 =  c \;
 (-1)^{\nu_1} \Gamma(m-\eps-\nu_{125}) \Gamma(1-m+\eps+\nu_{125})
}
\nonumber \\
 & & 
 \times
 \sum\limits_{n=0}^\infty
 \sum\limits_{j=0}^\infty
 \frac{\Gamma(n+m-\eps-\nu_4) \Gamma(n+2m-2\eps-\nu_{345})}
      {\Gamma(n+1) \Gamma(n+m-\eps)}
 \frac{\Gamma(j-m+\eps+\nu_{145}) \Gamma(j-2m+2\eps+\nu_{12345})}
      {\Gamma(j+1) \Gamma(j+m-\eps)}
 \nonumber \\
 & & 
 \times
 \frac{\Gamma(n+j+1) \Gamma(n+j+m-\eps)}
      {\Gamma(n+j+1+\nu_1) \Gamma(n+j+1-m+\eps+\nu_{125})}
 \nonumber \\ 
 & & 
 +
 c \;
 \frac{\Gamma(m-\eps-\nu_{14}) \Gamma(1-m+\eps+\nu_{14})
       \Gamma(2m-2\eps-\nu_{1345}) \Gamma(1-2m+2\eps+\nu_{1345})}
      {\Gamma(m-\eps-\nu_1) \Gamma(1-m+\eps+\nu_1)}
 \nonumber \\
 & & 
 \times
 \sum\limits_{j=0}^{\nu_1-1}
 \frac{\Gamma(-j+m-\eps-\nu_{25}) \Gamma(j+m-\eps-\nu_1)}
      {\Gamma(j+1) \Gamma(-j+\nu_1)}
 \sum\limits_{n=0}^\infty
 \frac{\Gamma(n-m+\eps+\nu_{145}) \Gamma(n-2m+2\eps+\nu_{12345})}
      {\Gamma(n+1) \Gamma(n+m-\eps)}
 \nonumber \\
 & & 
 \times
 \frac{\Gamma(n-j+\nu_1) \Gamma(n-j+1-m+\eps+\nu_1)}
      {\Gamma(n-j+1-m+\eps+\nu_{14}) \Gamma(n-j+1-2m+2\eps+\nu_{1345})}.
\eq
Once again, the sums are of the same type as discussed in eq. (\ref{T11noint}) and 
eq. (\ref{T11oneint}), and the Laurent expansion contains only rational
numbers and multiple zeta values.

To summarize, we are able to write the two-loop integral $\hat{I}^{(2,5)}$ 
as a combination of terms, which can
be expanded in $\eps$ to arbitrary order 
with the help of algorithms A and B of \cite{Moch:2001zr,Weinzierl:2002hv}.
All these terms occur with unit arguments, therefore the Laurent expansion of $\hat{I}^{(2,5)}$
involves only rational numbers and multiple zeta values.
In particular, in the representation derived here, no alternating Euler sums
occur.
This answers a question raised recently by Broadhurst \cite{Broadhurst:2002gb}:
\begin{theorem}
\label{TheBierenbaumWeinzierlTheorem}
Multiple zeta values are sufficient for the Laurent expansion of the two-loop integral
$\hat{I}^{(2,5)}(m-\eps,\nu_1,\nu_2,\nu_3,\nu_4,\nu_5)$, if all powers of 
the propagators are of the form $\nu_j=n_j+a_j\eps$, 
where the $n_j$ are positive integers and the $a_j$ are non-negative real numbers.
\end{theorem}
We have shown above, that the Laurent expansion of the two-loop integral 
$\hat{I}^{(2,5)}$
can be expressed in multiple zeta values if the 
inequalities eq. (\ref{ineq}) are satisfied.
These inequalities ensure that the semi-circles at infinity 
give a vanishing contribution.
If for a specific combination of powers of propagators these inequalities
are not satisfied, one may use the integration-by-part identities
eq. (\ref{ibp})
and the symmetry relations eq. (\ref{obvioussymmetry})
and eq. (\ref{sixcycle}) - (\ref{reflection})
to express the original integral as a linear combination of integrals, which fullfill
the inequalities.
The coefficients of this linear combination are either rational numbers or 
Gamma functions, which in turn expand into zeta values.
Therefore the theorem holds also in the general case.

The restriction to $a_j \ge 0$ in the parameterization 
$\nu_j=n_j+a_j\eps$ ensures that the different cases we treated above
(e.g. whether the argument of a Gamma function is an integer or not)
are sufficient. This condition could be relaxed at the expense of a more
extensive case study. However, in all practical calculations the integrals
occur with $a_j \ge 0$.

Finally we remark on a technical detail: 
Our method of calculation expresses the two-loop integral as a combination
of several sums with unit arguments. 
Although the final result is finite, it is not guaranteed 
that 
all individual sums are convergent.
Furthermore, the algorithms used
for the Laurent expansion rely on partial fractioning.
This may split a convergent sum into two divergent pieces, as
illustrated by the example of the convergent sum
\bq
\sum\limits_{n=1}^\infty \frac{1}{n(n+1)} & = & 1.
\eq
Partial fractioning splits
\bq
\frac{1}{n(n+1)} & = & \frac{1}{n} - \frac{1}{n+1}.
\eq
The problem is easily circumvented by introducing a finite upper summation
limit $N$ or by multiplying the summand by $x^n$ and taking the limit
$x \rightarrow 1$ in the end.
With both methods, the divergent pieces cancel at the end of the day
and one obtains the correct and finite result.
Within the second method, one inserts
\bq
 x_1^\sigma \; x_2^\tau
\eq
into the integrand of eq. (\ref{tobesolved}).
Then all sums are convergent, provided $0\le x_1<1$, $0 \le x_2<1$ and $x_2<x_1$.
The Laurent expansion for this more general expression will contain
multiple polylogarithms in $x_1$ and $x_2$.
The result for the two-loop integral $\hat{I}^{(2,5)}$
is recovered by first taking the limit $x_1 \rightarrow 1$
and then the limit $x_2 \rightarrow 1$.
Note that the order of the limits cannot be exchanged.
It is easily seen that in the limits the multiple polylogarithms reduce
to multiple zeta values.

\section{Results, checks and performance}
\label{sect:checks}

We have implemented the results of the previous section into a C++
program, which calculates symbolically the Laurent expansion of the two-loop
integral for a user-specified set of parameters $(m,\nu_1,...,\nu_5)$
up to the desired order in $\eps$.
The program uses the ``nestedsums''-library \cite{Weinzierl:2002hv}
and the ``GiNaC''-library \cite{Bauer:2000cp}.
To simplify our results we use the Gr\"obner basis for multiple zeta values 
provided in \cite{Minh}.

To check our implementation we have written two independent programs and verified that they agree.
Further we have checked that for trivial cases (e.g. all powers of the propagators integers)
we obtain the correct (known) result.
In addition, we have verified the symmetry relations eq.(\ref{obvioussymmetry}).
Finally we compared for non-integer powers of the propagators with known results
from the literature.
As an example we quote the result for one particular integral:
\bq
\lefteqn{
\left( 1-2\eps \right)
\hat{I}^{(2,5)}(2-\eps,1+\eps,1+\eps,1+\eps,1+\eps,1+\eps)
 = 
 6 \zeta_3 
 + 9 \zeta_4 \eps
 + 372 \zeta_5 \eps^2
}
\nonumber \\
 & &
 +\left(915 \zeta_6 -864 \zeta_3^2 \right) \eps^3
 +\left( 18450 \zeta_7 -2592 \zeta_4 \zeta_3 \right) \eps^4
 +\left(50259 \zeta_8-76680 \zeta_5 \zeta_3 -2592 \zeta_{6,2}\right) \eps^5
 \nonumber \\
 & &
 +\left(905368 \zeta_9 -200340 \zeta_6 \zeta_3 -130572 \zeta_5 \zeta_4  
        +66384 \zeta_3^3 \right) \eps^6
 \nonumber \\
 & &
 +\left(
        2955330\zeta_{10}
        -68688\zeta_{8,2}
        -3659904\zeta_7\zeta_3
        -1777680\zeta_5^2
        +298728\zeta_4\zeta_3^2
  \right) \eps^7
 + {\cal O}(\eps^8).
 \nonumber \\
\eq
For multiple zeta values we use the notation
\bq
\zeta_{m_1, m_2} & = & \sum\limits_{n_1>n_2>0}^\infty 
 \frac{1}{n_1^{m_1}}
 \frac{1}{n_2^{m_2}}.
\eq
We have compared our result up to weight 10 with \cite{Broadhurst:1997ur} and found agreement. 
The order to which we can expand the two-loop integral in the parameter $\eps$ is not
limited by our method of calculation.
The only restriction arises from hardware constraints (e.g. available memory and CPU time).
\begin{table}
\begin{center}
\begin{tabular}{|c|rrrrrrrr|} \hline
$O(\eps)$ & 0 & 1 & 2 & 3 & 4 & 5 & 6 & 7 \\
weight    & 3 & 4 & 5 & 6 & 7 & 8 & 9 & 10 \\ 
\hline 
time      & 29 & 54 & 99 & 185 & 375 & 910 & 2997 & 11741 \\
memory    & 6 & 7 & 8 & 12 & 30 & 104 & 397 & 1970 \\
\hline
\end{tabular}
\end{center}
\caption{\label{bench}
\it CPU time in seconds and required memory in MB 
for the expansion of 
$\hat{I}^{(2,5)}(2-\eps,1+\eps,1+\eps,1+\eps,1+\eps,1+\eps)$ 
up to the indicated order/weight on a PC (1.6 GHz Athlon with 2 GB RAM).}
\end{table}
Table (\ref{bench}) shows 
for the case of the integral
$\hat{I}^{(2,5)}(2-\eps,1+\eps,1+\eps,1+\eps,1+\eps,1+\eps)$ 
the dependence of the required CPU time and the required memory
on the order of $\eps$ to which the Laurent expansion is calculated.
In table (\ref{bench}) we also indicate the highest weight of multiple zeta values,
which occur within a given order.
Note that the weight is a more accurate measure of the complexity of a calculation,
since in the example discussed here individual terms start at $1/\eps^3$, but in the
sum the coefficients of the terms $1/\eps^3$, $1/\eps^2$ and $1/\eps$ vanish.
In general we expect terms up to weight $2l-1$ to occur in the finite
part of a $l$-loop two-point function.
We see from table (\ref{bench}) that up to weight 8 our implementation 
is rather efficient.
For higher weights the main limitation is given by the available memory.
However, as far as practical applications are concerned, weight 7 is sufficient
to extend the exisiting ${\cal O}(\alpha_s^3)$ calculation \cite{Gorishnii:1991vf,Surguladze:1991tg}
for $e^+ e^- \rightarrow \mbox{hadrons}$ 
to order $\alpha_s^4$.
Here one would need the following integrals:
\bq
\lefteqn{
\left( 1-2\eps \right)
\hat{I}^{(2,5)}(2-\eps,1+2\eps,1,1,1,1)
 = 
 6 \zeta_3+9 \zeta_4 \eps +142 \zeta_5 \eps^2
 +\left( 340 \zeta_6 -158 \zeta_3^2 \right) \eps^3
 }
 \nonumber \\
 & &
 +\left( 3034 \zeta_7 -474 \zeta_4 \zeta_3 \right) \eps^4
 +\left( \frac{36099}{4} \zeta_8 - 6172 \zeta_5 \zeta_3 \right) \eps^5
\hspace*{5cm}
 \nonumber \\ 
 & &
 +\left( \frac{193010}{3} \zeta_9 -9258 \zeta_5 \zeta_4 
         -14640 \zeta_6 \zeta_3 + \frac{6748}{3} \zeta_3^3 
  \right) \eps^6
 + {\cal O}(\eps^7),
 \nonumber \\
\lefteqn{
\left( 1-2\eps \right)
\hat{I}^{(2,5)}(2-\eps,1,1,1,1,1+2\eps)
 = 
 6 \zeta_3+9 \zeta_4 \eps+192 \zeta_5 \eps^2
 +\left(465 \zeta_6-168 \zeta_3^2 \right) \eps^3
 }
 \nonumber \\
 & &
 +\left(4509 \zeta_7-504 \zeta_4 \zeta_3 \right) \eps^4
 +\left(\frac{16377}{2} \zeta_8 -1620 \zeta_{6,2}-3252 \zeta_5 \zeta_3  \right) \eps^5
 \nonumber \\ 
 & &
 +\left(98490 \zeta_9-14598 \zeta_5 \zeta_4-15390 \zeta_6 \zeta_3  +2676\zeta_3^3 \right) \eps^6
 + {\cal O}(\eps^7),
 \nonumber \\
\lefteqn{
\left( 1-2\eps \right)
\hat{I}^{(2,5)}(2-\eps,1+\eps,1+\eps,1,1,1)
 = 
 6 \zeta_3+9 \zeta_4 \eps+132\zeta_5 \eps^2
 +\left(315 \zeta_6-144 \zeta_3^2 \right) \eps^3
 }
 \nonumber \\
 & &
 +\left(2634 \zeta_7-432 \zeta_4 \zeta_3 \right) \eps^4
 +\left(7749 \zeta_8-5256 \zeta_5\zeta_3  \right) \eps^5
 \nonumber \\
 & &
 +\left(53160 \zeta_9-12420 \zeta_6 \zeta_3-7884 \zeta_5\zeta_4  +1872 \zeta_3^3\right) \eps^6
 + {\cal O}(\eps^7),
 \nonumber \\
\lefteqn{
\left( 1-2\eps \right)
\hat{I}^{(2,5)}(2-\eps,1+\eps,1,1+\eps,1,1)
 = 
 6\zeta_3+9 \zeta_4 \eps+127\zeta_5\eps^2
 +\left(\frac{605}{2}\zeta_6-173\zeta_3^2\right) \eps^3
 }
 \nonumber \\
 & &
 +\left(\frac{18989}{8}\zeta_7-519\zeta_4 \zeta_3 \right) \eps^4
 +\left(\frac{102243}{16}\zeta_8-\frac{243}{2}\zeta_{6,2}-5839 \zeta_5\zeta_3 \right) \eps^5
 \nonumber \\
 & &
 +\left(\frac{1084927}{24}\zeta_9-14340 \zeta_6 \zeta_3-\frac{18975}{2}\zeta_5 \zeta_4 +\frac{8554}{3}\zeta_3^3 \right)\eps^6
 + {\cal O}(\eps^7),
 \nonumber \\
\lefteqn{
\left( 1-2\eps \right)
\hat{I}^{(2,5)}(2-\eps,1+\eps,1,1,1+\eps,1)
 = 
 6\zeta_3+9\zeta_4\eps+132\zeta_5\eps^2
 +\left( 315\zeta_6-204\zeta_3^2 \right)\eps^3
 }
 \nonumber \\
 & &
 +\left(2634\zeta_7-612\zeta_4\zeta_3\right) \eps^4
 +\left(\frac{15183}{2}\zeta_8-7476\zeta_5\zeta_3 \right) \eps^5
 \nonumber \\
 & &
 +\left(53160 \zeta_9-17670 \zeta_6 \zeta_3-11214 \zeta_5\zeta_4 +3612 \zeta_3^3 \right)\eps^6
 + {\cal O}(\eps^7),
 \nonumber \\
\lefteqn{
\left( 1-2\eps \right)
\hat{I}^{(2,5)}(2-\eps,1+\eps,1,1,1,1+\eps)
 = 
 6\zeta_3+9\zeta_4 \eps+157\zeta_5\eps^2
 +\left(\frac{755}{2}\zeta_6-179\zeta_3^2 \right) \eps^3
 }
 \nonumber \\
 & &
 +\left(\frac{26657}{8}\zeta_7-537\zeta_4\zeta_3 \right) \eps^4
 +\left(\frac{124899}{16}\zeta_8-\frac{1215}{2}\zeta_{6,2}-5521\zeta_5\zeta_3  \right)\eps^5
 \nonumber \\
 & &
 +\left(\frac{1657525}{24}\zeta_9-15945\zeta_6\zeta_3-\frac{23853}{2}\zeta_5\zeta_4+\frac{8776}{3}\zeta_3^3 \right)\eps^6
 + {\cal O}(\eps^7).
\eq
These integrals are in principle only needed up to order $\eps^4$ for the 
$\alpha_s^4$ corrections. However, knowing these integrals to higher orders
can simplify integral reduction algorithms.
For example, one may allow spurious poles in integration-by-parts identities.
The CPU time and the required memory needed to calculate each of these
integrals is slightly below the corresponding 
numbers indicated in table (\ref{bench}).

\section{Summary and conclusions}
\label{sect:concl}

In this paper we calculated the $\eps$-expansion of
the massless master two-loop two-point function 
with arbitrary powers of the propagators.
We showed that to all orders in $\eps$ rational numbers and multiple zeta values
are sufficient to express the result.
Our method of calculation obtained the two-loop integral from a convolution of two
one-loop integrals.
We also discussed the corresponding factorization for three-loop two-point
functions.
Finally we demonstrated that our method can be implemented efficiently on a computer.

\subsection*{Acknowledgements}
\label{sec:acknowledgements}

We would like to thank Dirk Kreimer and David Broadhurst
for useful discussions and comments on the manuscript.
S.W. would like to thank also Sven Moch for useful discussions.
I.B. acknowledges support by the Graduiertenkolleg 
``Eichtheorien - experimentelle Tests und theoretische Grundlagen'' 
at Mainz University.

\begin{appendix}

\section{Integral transformations}
\label{sect:trafo}

In this appendix we briefly summarize the Laplace, Mellin and Fourier integral transformations.
Let $f(t)$ be a function which is bounded by an exponential function for $t \rightarrow \pm \infty$, e.g.
\bq
\left| f(t) \right| \le K e^{c_0 t} & & \mbox{for}\;\; t \rightarrow \infty,
 \nonumber \\
\left| f(t) \right| \le K' e^{-c_1 t} & & \mbox{for}\;\; t \rightarrow -\infty.
\eq
Then the (double-sided) Laplace transform is defined for
$c_0 < \mbox{Re}\; \sigma < c_1$
by
\bq
f_{\cal L}(\sigma) & = &
 \int\limits_{-\infty}^\infty dt \; f(t) \; e^{-\sigma t}.
\eq
The inverse Laplace transform is given by
\bq
\label{inverselaplace}
f(t) & = & \frac{1}{2\pi i} \int\limits_{\gamma-i\infty}^{\gamma+i\infty}
 d\sigma \; f_{\cal L}(\sigma) \; e^{\sigma t}.
\eq
The integration contour is parallel to the imaginary axis and $c_0 < \mbox{Re}\; \gamma < c_1$.
\\
Let $h(x)$ be a function which is bounded by a power law for $x\rightarrow 0$ and $x \rightarrow \infty$,
e.g.
\bq
\left| h(x) \right| \le K x^{-c_0} & & \mbox{for}\;\; x \rightarrow 0,
 \nonumber \\
\left| h(x) \right| \le K' x^{c_1} & & \mbox{for}\;\; x \rightarrow \infty.
\eq
Then the Mellin transform is defined for
$c_0 < \mbox{Re}\; \sigma < c_1$
by
\bq
h_{\cal M}(\sigma) & = &
 \int\limits_{0}^\infty dx \; h(x) \; x^{\sigma-1}.
\eq
The inverse Mellin transform is given by
\bq
\label{inversemellin}
h(x) & = & \frac{1}{2\pi i} \int\limits_{\gamma-i\infty}^{\gamma+i\infty}
 d\sigma \; h_{\cal M}(\sigma) \; x^{-\sigma}.
\eq
The integration contour is parallel to the imaginary axis and $c_0 < \mbox{Re}\; \gamma < c_1$.
There is a close relation between the Laplace and the Mellin transform: 
If $f(t) = h(e^{-t})$, then $f_{\cal L}(\sigma) = h_{\cal M}(\sigma)$.
Note that the inversion formulas can be obtained from the known inversion formula for
the Fourier transform. To a function $f(t)$ we associate the Fourier transform
\bq
f_{\cal F}(u) & = & 
 \int\limits_{-\infty}^\infty dt \; f(t) \; e^{-2\pi i u t}.
\eq
The inverse transform is given by
\bq
f(t) & = &
 \int\limits_{-\infty}^\infty du \; f_{\cal F}(u) \; e^{2\pi i u t}.
\eq
The Laplace transform and the Fourier transform of a function $f(t)$ are related by
\bq
f_{\cal L}\left( \sigma \right) & = & f_{\cal F}\left( \frac{\sigma}{2 \pi i} \right).
\eq
As an example for the Mellin transform we consider the function 
\bq
h(x) & = & \frac{x^c}{(1+x)^c}
\eq
with Mellin transform $h_{\cal M}(\sigma)=\Gamma(-\sigma) \Gamma(\sigma+c) / \Gamma(c)$.
For $\mbox{Re}(-c) < \mbox{Re} \; \gamma < 0$ we have
\bq
\label{baseMellin}
\frac{x^c}{(1+x)^c}
 & = & 
\frac{1}{2\pi i} \int\limits_{\gamma-i\infty}^{\gamma+i\infty}
 d\sigma \; \frac{\Gamma(-\sigma) \Gamma(\sigma+c)}{\Gamma(c)} \; x^{-\sigma}.
\eq
From eq. (\ref{baseMellin}) one obtains with $x=B/A$ the Mellin-Barnes formula
\bq
\left(A+B\right)^{-c}
 & = & 
\frac{1}{2\pi i} \int\limits_{\gamma-i\infty}^{\gamma+i\infty}
 d\sigma \; \frac{\Gamma(-\sigma) \Gamma(\sigma+c)}{\Gamma(c)} \; A^\sigma B^{-\sigma-c}.
\eq
We often deal with integrals of the form
\bq
\label{MellinBarnesInt}
I
 & = & 
\frac{1}{2\pi i} \int\limits_{\gamma-i\infty}^{\gamma+i\infty}
 d\sigma \; 
 \frac{\Gamma(\sigma+a_1) ... \Gamma(\sigma+a_m)}
      {\Gamma(\sigma+c_2) ... \Gamma(\sigma+c_p)}
 \frac{\Gamma(-\sigma+b_1) ... \Gamma(-\sigma+b_n)}
      {\Gamma(-\sigma+d_1) ... \Gamma(-\sigma+d_q)} 
 \; x^{-\sigma}.
\eq
If $\;\mbox{max}\left( \mbox{Re}(-a_1), ..., \mbox{Re}(-a_m) \right) < \mbox{min}\left( \mbox{Re}(b_1), ..., \mbox{Re}(b_n) \right)$ the contour can be chosen
as a straight line parallel to the imaginary axis with
\bq
\mbox{max}\left( \mbox{Re}(-a_1), ..., \mbox{Re}(-a_m) \right) 
 \;\;\; < \;\;\; \mbox{Re} \gamma \;\;\; < \;\;\;
\mbox{min}\left( \mbox{Re}(b_1), ..., \mbox{Re}(b_n) \right),
\eq
otherwise the contour is indented, such that the residues of
$\Gamma(\sigma+a_1)$, ..., $\Gamma(\sigma+a_m)$ are to the right of the contour,
whereas the residues of 
$\Gamma(-\sigma+b_1)$,  ..., $\Gamma(-\sigma+b_n)$ are to the left of the contour.
We further set
\bq
\alpha & = & m+n-p-q,
\nonumber \\
\beta & = & m-n-p+q, 
\nonumber \\
\lambda & = & \mbox{Re} \left( \sum\limits_{j=1}^m a_j
                              +\sum\limits_{j=1}^n b_j
                              -\sum\limits_{j=1}^p c_j
                              -\sum\limits_{j=1}^q d_j \right)
              - \frac{1}{2} \left( m+n-p-q \right).
\eq
Then the integral eq. (\ref{MellinBarnesInt})
converges absolutely for $\alpha >0$ \cite{Erdelyi} and defines an analytic function in
\bq
\left| \mbox{arg} \; x \right| & < & \mbox{min}\left( \pi, \alpha \frac{\pi}{2} \right).
\eq
The integral eq. (\ref{MellinBarnesInt}) is most conveniently evaluated with 
the help of the residuum theorem by closing the contour to the left or to the right.
Therefore we need to know under which conditions the semi-circle at inifinty used to close the contour gives a vanishing contribution.
This is obviously the case for $|x|<1$ if we close the contour to the left,
and for $|x|>1$, if we close the contour to the right.
The case $|x|=1$ deserves some special attention. One can show that
in the case $\beta=0$ the semi-circle gives a vanishing contribution, provided
\bq
\lambda & < & -1.
\eq
To derive this result, Barnes asymptotic expansion of the Gamma function for large $x$ is 
useful:
\bq
\ln \Gamma(x+c) & \sim & 
(x+c) \ln x  - x - \frac{1}{2} \ln \frac{x}{2\pi}
 - \sum\limits_{n=1}^\infty \frac{B_{n+1}(c)}{n(n+1)} \left( - \frac{1}{x} \right)^n,
\eq
where the Bernoulli polynomials $B_n(x)$ 
are given in terms of the Bernoulli numbers $B_j$ by
\bq
B_n(x) & = & \sum\limits_{j=0}^n
\left( \begin{array}{c} n \\ j \\ \end{array} \right)
B_j \; x^{n-j}.
\eq

\end{appendix}



\begin{thebibliography}{10}

\bibitem{Gorishnii:1989gt}
S.~G. Gorishnii, S.~A. Larin, L.~R. Surguladze, and F.~V. Tkachov,
\newblock Comput. Phys. Commun. {\bf 55}, 381 (1989).

\bibitem{Larin:1991fz}
S.~A. Larin, F.~V. Tkachov, and J.~A.~M. Vermaseren,
\newblock NIKHEF-H-91-18.

\bibitem{Chetyrkin:1980pr}
K.~G. Chetyrkin, A.~L. Kataev, and F.~V. Tkachov,
\newblock Nucl. Phys. {\bf B174}, 345 (1980).

\bibitem{Kazakov:1983km}
D.~I. Kazakov,
\newblock Phys. Lett. {\bf B133}, 406 (1983).

\bibitem{Kazakov:1984ns}
D.~I. Kazakov,
\newblock Theor. Math. Phys. {\bf 58}, 223 (1984).

\bibitem{Kazakov:1985pk}
D.~I. Kazakov,
\newblock Theor. Math. Phys. {\bf 62}, 84 (1985).

\bibitem{Broadhurst:1986bx}
D.~J. Broadhurst,
\newblock Z. Phys. {\bf C32}, 249 (1986).

\bibitem{Barfoot:1988kg}
D.~T. Barfoot and D.~J. Broadhurst,
\newblock Z. Phys. {\bf C41}, 81 (1988).

\bibitem{Kotikov:1996cw}
A.~V. Kotikov,
\newblock Phys. Lett. {\bf B375}, 240 (1996), hep-ph/9512270.

\bibitem{Broadhurst:1997ur}
D.~J. Broadhurst, J.~A. Gracey, and D.~Kreimer,
\newblock Z. Phys. {\bf C75}, 559 (1997), hep-th/9607174.

\bibitem{Broadhurst:2002gb}
D.~J. Broadhurst,
\newblock Nucl. Phys. Proc. Suppl. {\bf 116}, 432 (2003), hep-ph/0211194.

\bibitem{Gorishnii:1985te}
S.~G. Gorishnii and A.~P. Isaev,
\newblock Theor. Math. Phys. {\bf 62}, 232 (1985).

\bibitem{Smirnov:2002kq}
V.~A. Smirnov,
\newblock (2002), hep-ph/0209177.

\bibitem{Weinzierl:2003jx}
S.~Weinzierl,
\newblock (2003), hep-th/0305260.

\bibitem{Grozin:2003ak}
A.~G. Grozin,
\newblock (2003), hep-ph/0307297.

\bibitem{Moch:2001zr}
S.~Moch, P.~Uwer, and S.~Weinzierl,
\newblock J. Math. Phys. {\bf 43}, 3363 (2002), hep-ph/0110083.

\bibitem{Weinzierl:2002hv}
S.~Weinzierl,
\newblock Comput. Phys. Commun. {\bf 145}, 357 (2002), math-ph/0201011.

\bibitem{'tHooft:1972fi}
G.~'t~Hooft and M.~J.~G. Veltman,
\newblock Nucl. Phys. {\bf B44}, 189 (1972).

\bibitem{Tkachov:1981wb}
F.~V. Tkachov,
\newblock Phys. Lett. {\bf B100}, 65 (1981).

\bibitem{Chetyrkin:1981qh}
K.~G. Chetyrkin and F.~V. Tkachov,
\newblock Nucl. Phys. {\bf B192}, 159 (1981).

\bibitem{vanRitbergen:1999fi}
T.~van Ritbergen and R.~G. Stuart,
\newblock Nucl. Phys. {\bf B564}, 343 (2000), hep-ph/9904240.

\bibitem{Kreimer:2002qy}
D.~Kreimer,
\newblock Ann. Phys. {\bf 303}, 179 (2003), hep-th/0211136.

\bibitem{Kreimer:2002fv}
D.~Kreimer,
\newblock Nucl. Phys. Proc. Suppl. {\bf 116}, 392 (2003), hep-ph/0211188.

\bibitem{Kreimer:2003vp}
D.~Kreimer,
\newblock (2003), hep-th/0306020.

\bibitem{Smirnov:2003vi}
V.~A. Smirnov,
\newblock Phys. Lett. {\bf B567}, 193 (2003), hep-ph/0305142.

\bibitem{Bauer:2000cp}
C.~Bauer, A.~Frink, and R.~Kreckel,
\newblock J. Symbolic Computation {\bf 33}, 1 (2002), cs.sc/0004015.

\bibitem{Vermaseren:1998uu}
J.~A.~M. Vermaseren,
\newblock Int. J. Mod. Phys. {\bf A14}, 2037 (1999), hep-ph/9806280.

\bibitem{Remiddi:1999ew}
E.~Remiddi and J.~A.~M. Vermaseren,
\newblock Int. J. Mod. Phys. {\bf A15}, 725 (2000), hep-ph/9905237.

\bibitem{Blumlein:1998if}
J.~Bl\"umlein and S.~Kurth,
\newblock Phys. Rev. {\bf D60}, 014018 (1999), hep-ph/9810241.

\bibitem{Minh}
H.~N. Minh and M.~Petitot, 
\newblock Discrete Math. {\bf 217}, 273 (2000).

\bibitem{Gorishnii:1991vf}
S.~G. Gorishnii, A.~L. Kataev, and S.~A. Larin,
\newblock Phys. Lett. {\bf B259}, 144 (1991).

\bibitem{Surguladze:1991tg}
L.~R. Surguladze and M.~A. Samuel,
\newblock Phys. Rev. Lett. {\bf 66}, 560 (1991), Erratum ibid. {\bf 66}, 2416 (1991).

\bibitem{Erdelyi}
A.~ Erd{\'e}lyi et~al.,
\newblock  "Higher Transcendental Functions", (Vol. I, McGraw Hill, 1953).

\end{thebibliography}
\end{document}